\documentclass[openacc]{rsproca_new}

\usepackage{graphicx}
\usepackage{amsmath}
\usepackage{amssymb}
\usepackage{natbib}
\usepackage{lineno}
\setcitestyle{numbers,open={[},close={]},citesep={,}}

\begin{document}

\title{Rheology of three-phase suspensions determined via dam-break experiments}

\author{J. Birnbaum$^{1}$, E. Lev$^{1}$ and  E. W. Llewellin$^{2}$}

\address{$^{1}$Lamont-Doherty Earth Observatory, Columbia University, 61 Rte. 9w, Palisades, NY, 10964, US \\
$^{2}$Department of Earth Sciences, Durham University, Durham DH1 3LE, UK}

\subject{fluid mechanics, material science, volcanology}


\corres{J. Birnbaum\\
\email{janineb@ldeo.columbia.edu}}

\begin{abstract}
Three-phase suspensions, of liquid that suspends dispersed solid particles and gas bubbles, are common in both natural and industrial settings. Their rheology is poorly constrained, particularly for high total suspended fractions ($\gtrsim 0.5$). We use a dam-break consistometer to characterize the rheology of suspensions of (Newtonian) corn syrup, plastic particles, and CO$_2$ bubbles. The study is motivated by a desire to understand the rheology of magma and lava. Our experiments are scaled to the volcanic system: they are conducted in the non-Brownian, non-inertial regime; bubble capillary number is varied across unity; and bubble and particle fractions are $0\leq\phi_{gas}\leq0.82$ and $0\leq\phi_{solid}\leq0.37$ respectively. We measure flow-front velocity and invert for a Herschel--Bulkley rheology model as a function of $\phi_{gas}$, $\phi_{solid}$, and the capillary number. We find a stronger increase in relative viscosity with increasing $\phi_{gas}$ in the low to intermediate capillary number regime than predicted by existing theory, and find both shear-thinning and shear-thickening effects, depending on the capillary number. We apply our model to the existing community code for lava flow emplacement, PyFLOWGO, and predict increased viscosity and decreased velocity compared with current rheological models, suggesting existing models may not adequately account for the role of bubbles in stiffening lavas.
\end{abstract}


\begin{fmtext}

\section{Introduction}
Two-phase liquid--solid (particle) and liquid--gas (bubble) suspensions are important across nature and industry. Consequently, their behavior has been well characterized: see \citet{Mader2013} for a through review. Numerous stu-

\end{fmtext}


\maketitle

\noindent
dies show that the apparent viscosity of particle suspensions increases with particle volume fraction, and intermediate and concentrated suspensions show non-Newtonian behavior with finite yield stresses and strain-rate dependence \citep{Einstein1911,Krieger1959,Mueller2010}. Bubble suspensions exhibit an increase or decrease in apparent viscosity relative to the liquid phase. Concentrated bubble suspensions also show non-Newtonian behavior, including non-zero yield stresses and shear-thinning or shear-thickening behavior \citep{Chesterton2013, Llewellin2002a, Llewellin2005, Manga1998, Pal2003, Pistone2012, Princen1989, Rouyer2005, Rust2002, Stein1992, Taylor1932}. A key parameter determining the rheology of bubble suspensions is the capillary number: 
\begin{equation}
    \mathrm{Ca} = \frac{\mu a \dot{\gamma}}{\Gamma} \: , 
    \label{eq:Ca}
\end{equation}
where $\mu$ is the liquid viscosity, $a$ is the undeformed bubble radius, $\dot{\gamma}$ is the shear strain rate, and $\Gamma$ is the surface tension. The capillary number reflects the ratio of viscous stresses that deform bubbles and capillary stresses that restore them \citep{Stein1992, Rust2002, Llewellin2005} When $\mathrm{Ca} \gg 1$, viscous stress dominates, bubbles deform easily and decrease relative viscosity; when $\mathrm{Ca} \ll 1$, the surface tension resisting bubble deformation dominates, bubbles remain nearly spherical and increase relative viscosity. \par

The rheology of three-phase suspensions (liquid suspending particles and bubbles) has received much less detailed investigation, but is known to be a complex function of the suspended phase fractions, micro-textural properties such the size and shape distributions of particles and bubbles, and the conditions of shear \citep{Phan-Thien1997,Pistone2012,Truby2015}. The only systematic experimental study, which was limited to $\phi_{solid} \lesssim 0.5$, $\phi_{gas} \lesssim 0.3$ and low capillary number \citep{Truby2015}, found that the rheology was well described by a simple convolution of existing two-phase rheology models in which the liquid and bubbles were treated as an effective medium that suspended the particles. We expect that interactions between bubbles and particles will become increasingly important at higher bubble and particle fractions such that a simple convolution of two-phase models is insufficient. 

Magma (and lava, which is its subaerial counterpart) is a natural three-phase suspension, composed of a molten silicate liquid (melt) that suspends a variable fraction of solid particles (crystals) and gas bubbles. The suspending melt is Newtonian over a wide range of strain rates, and has a viscosity that can vary over orders of magnitude \citep{Dingwell1989,Giordano2008}. Bubble and crystal volume fractions can both range from 0 to 1, and typically change, along with melt viscosity, during transport through the Earth's crust and over its surface. The rheology of magma exerts a first-order control on the dynamics of its eruption, and the emplacement of subsequent lava flows \citep{gonnermann2007,Cashman2013}. Whilst the rheology of natural magmas has been measured directly \citep{Ryerson1988, Pinkerton1995, Ishibashi2009, Soldati2016}, such measurements are challenging to perform and interpret, and are not well suited to systematic investigation of parameter space. As a result, it has been common for rheology of magmatic suspensions to be investigated via analogue experiments \citep{Llewellin2002a,Rust2002,Mueller2010,Castruccio2010,Truby2015}. Even with the use of analogues, it has proven challenging to perform rheometry on samples with $\phi_{gas} \gtrsim 0.5$. This is because samples with high bubble fraction are difficult to prepare, particularly when the sample also contains particles, and because they are prone to breakdown when loaded into a conventional rheometer. We conduct experiments in which bubbles are grown \textit{in situ} via a chemical reaction, and analysed using a dam-break consistometer. This approach circumvents both problems, allowing us to investigate samples with bubble fractions ($0\leq\phi_{gas}\leq0.8$) and particle fractions ($0\leq\phi_{solid}\leq0.37$) that span the most relevant ranges for natural magma and lava. 

\section{Theoretical background}
\label{section:theoretical_framework}
\subsection{Scaling}
We scale our analogue suspensions to basaltic magma, which typically has a pure-melt viscosity in the range 10$^2$ to 10$^4$ Pas ($10^3-10^5$ Poise) at eruption temperature \citep{Giordano2008}, and surface tension between 0.05 and 0.3 N/m \citep{Bagdassarov2000, Mangan2005, Walker1981}. Bubbles that nucleate deep within the volcanic plumbing system are likely to be small (radius $\sim 10^{-5}$ to $10^{-3}$ m) and experience very low shear rates ($\sim 10^{-10}$ to $10^{-5}$ 1/s), resulting in capillary numbers $\mathrm{Ca}\ll1$ \citep{Petford2020, Vergniolle1996}. As magmas decompress on their way to the surface, bubbles grow and coalesce to radii $\sim 10^{-5} - 10^{-2}$ m or larger \citep{Galindo2012, Gaonach2005, Thivet2020}. However, decimeter sized bubbles often segregate from the flow, making a  suspension rheology approach no longer appropriate. Magma in basaltic dikes experiences strain rates up to $\sim 10^{1}$ to $10^{2}$ 1/s \citep{Petcovic2005} and capillary number therefore spans from $\mathrm{Ca}\ll1$ to $\mathrm{Ca}\gg1$. At the surface, lava flows, particularly fast-moving flows (velocities $\sim$1 m/s) may experience strain rates of $10^{-3}$ to $10^2$ 1/s, and both high and low capillary numbers \citep{Manga1998}. The viscosity of silicate melts is sufficiently high that viscous stresses at the particle scale dominate over inertial and Brownian stresses \citep{Mueller2010}. We therefore conduct experiments that span from low to high capillary number, whilst remaining in the non-inertial, non-Brownian regimes. \par

\subsection{Theoretical framework for rheology of three-phase suspensions}
To our knowledge, the only theoretically-grounded model for the rheology of three-phase suspensions is that of \citet{Phan-Thien1997}. They present a model for the effective Newtonian viscosity, $\eta$, of a three-phase suspension of the form:
\begin{equation}
    \eta_{r,s} = \frac{\eta}{\mu} = \eta_{r,p} \: \eta_{r,b} 
    \label{eq:theory_3phase_consistency}
\end{equation}
where $\eta_{r,s}$ is the relative viscosity of the three-phase suspension (i.e. suspension viscosity normalized by the liquid viscosity), and $\eta_{r,p}$ and $\eta_{r,b}$ are the relative viscosities of the two-phase components (particle-liquid, and bubble-liquid suspensions respectively). Importantly, this relation treats one two-phase system as an effective viscous medium in which the third phase is suspended. \citet{Truby2015} validated this theoretical form against experimental data, using existing models for the viscosity of the two-phases suspensions: 
\begin{align}
    \begin{split}
        \eta_{r,p} = \left( 1 - \frac{\phi_{solid}}{\phi_m} \right)^{-B_{solid}}, \qquad\mathrm{and}\qquad 
        \eta_{r,b} = \left( 1 - \phi_{gas} \right)^{-B_{gas}} \: ,
    \end{split}
\end{align}
following, respectively, \citet{Mueller2010} and \citet{Llewellin2005}, where Einstein exponents $B_{solid}=2$ and $B_{gas}=1$, and $\phi_m$ is a maximum packing fraction of solid particles. Both \citet{Phan-Thien1997} and \citet{Truby2015} note that the definition of $\phi_{gas}$ and $\phi_{solid}$ depends on which two-phase suspension is chosen as the effective medium; in this study we choose the particle suspension such that $\phi_{solid} = V_{solid}/(V_{solid} + V_{liquid})$ and $\phi_{gas} = V_{gas}/(V_{gas} + V_{solid} + V_{liquid})$, where $V$ denotes the volume of the subscript phase. \par 
\citet{Truby2015} extend the model of \citet{Phan-Thien1997} to allow for non-Newtonian effects by equating the suspension viscosity, $\eta$, with the consistency, $K$, in the Herschel-Bulkley model \citep{Herschel1926}:
\begin{equation}
    \tau = \tau_y + K \dot{\gamma}^n \: , 
    \label{eq:herschel-bulkley}
\end{equation}
where $\tau$ is the shear stress, $\tau_y$ is the yield stress, and $n$ is the flow index  ($n>1$ implies shear-thickening, $n<1$ shear-thinning). We adopt this approach in the analysis below.

\section{Methods}
\subsection{Experimental setup}
\begin{figure}[!htb]
\begin{center}
\includegraphics[width=\textwidth]{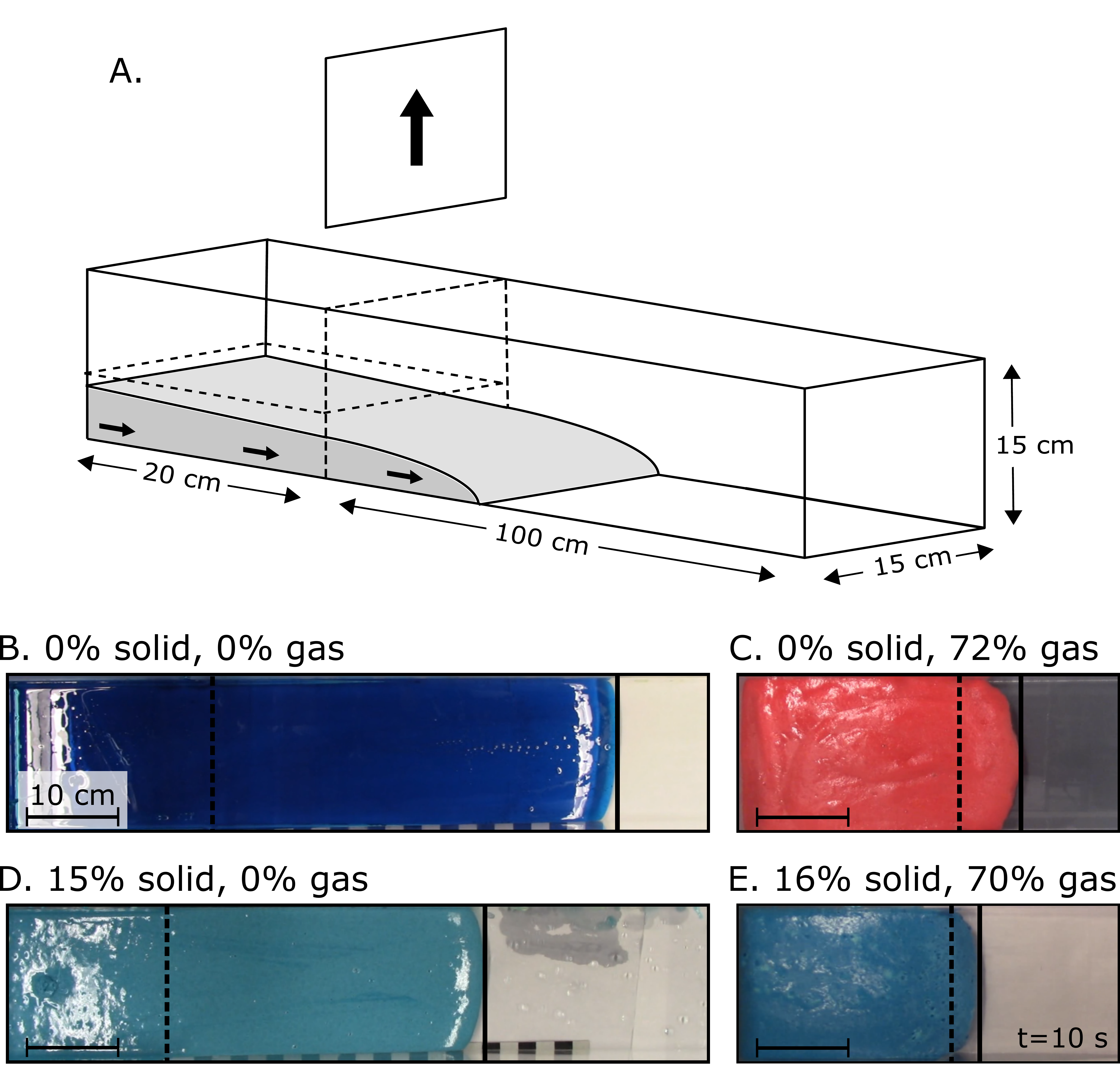}\\
\caption{A) Dam-break setup showing fluid release into a confined channel. B-E) Top views contrasting the slower flow-front progression with the addition of particles and bubbles, with starting position (dashed) and position 10 s after dam removal (solid). All photos have the same scale.} 
\label{fig:Exp_setup}
\end{center}
\end{figure}

Our experiments use a Bostwick (dam-break) consistometer, in which fluid is released from a reservoir and into a confined rectangular channel (Fig. \ref{fig:Exp_setup}). The reservoir measures 20 cm along the length of the channel, and the channel extends another 1 m; the reservoir and channel have a constant width of 15 cm. The reservoir can be filled to a maximum depth of 15 cm, and in our experiments initial reservoir depth ranges from 4 to 13 cm. The walls of the consistometer are transparent to allow observation and imaging of the flow. We use 42DE corn syrup diluted to a 77\% sugar concentration, measured using a Brix refractometer. Corn syrup is a Newtonian fluid with a viscosity that depends on water-content (dilution) and temperature. Our experiments are conducted at room temperature (isothermal for a given experiment) which varied between 18.1 and 30.5 $^\circ$C, corresponding to syrup viscosities in the range 13.3 $\leq \mu \leq$ 3.73 Pa s, measured using a Brookfield rotational viscometer. The surface tension of corn syrup is $\Gamma=0.08$ N/m \citep{Rust2002a}; the surface tension of sugar solutions is only weakly dependent on concentration \citep{Sinat-Radchenko}.\par 

To the syrup, we add rigid plastic particles sieved to a mesh size of 60-80 (Sourced from Precise Finishing Inc Stock No. GP-PP60/80, Polyplus / Type II). Particles have near-neutral buoyancy ($\rho_{syrup} = 1395$ kg/m$^3$, $\rho_{plastic} = 1370$ kg/m$^3$), are angular and slightly elongate (aspect ratio 1.5 to 2), and have a dominantly unimodal size distribution (long axis $\approx 200$ to $400\,\mu$m). Particles are stirred into the suspension by hand which leads to minimal air entrapment as shown in the microphotograph in Supplementary Fig. \ref{fig:Bubble_def}. Volume fraction of solids in the suspension is determined prior to the introduction of gas by mass of syrup and particles (uncertainty of 1-2 volume \%). When multiple experiments with the same particle content were performed, a large volume of syrup and particles was prepared, but gas was introduced to each sample individually. \par

Gas is introduced via a chemical reaction between baking soda (NaHCO$_3$) and citric acid (C$_6$H$_8$O$_7$) which produces water vapor and CO$_2$ gas. Reactants are added with a $\sim$3:4 ratio by mass, yielding an empirical relation of roughly 0.5 g baking soda and 0.7 g citric acid per 100 g syrup, per 10\% desired gas volume fraction. The reaction begins immediately and the sample is allowed to rise for $\sim$20 min prior to its release into the confined channel. Samples with $\phi_{gas} < 0.6$ are risen in a separate container, then transferred to the consistometer reservoir; samples with $\phi_{gas}>0.6$ are more delicate, and are risen in the reservoir to avoid foam disruption during transfer.  Final gas volume fraction is calculated from the final mixture density. Mass of the suspension is measured when transferring to the channel and the volume is calculated using the internal channel dimensions and the height of the mixture within the reservoir as measured through the channel walls. This method leads to $\sim$3 to 5\% relative uncertainty on the volume and resultant $\sim$3 to 5\% relative uncertainty on the density and an absolute uncertainty of $\sim$0.02 to 0.04 on $\phi_{gas}$. \par 

The water vapor that forms  during the reaction is insufficient to lower the syrup viscosity appreciably. In the experiments with the highest concentration of reactants (9 g total per 100 g of syrup), that water could, in principle, result in a dilution from 77\% to 76.6\% sugar if all the water dissolved, which would change the viscosity from 7.6 Pas (76 Poise) to 6.5 Pas (65 Poise) at 23$^\circ$C. However, a test suspension, from which the evolved gas was allowed to escape overnight showed no additional dissolved water when analysed using the Brix refractometer. \par

The flow begins with the removal of the dam, which takes less than one second. The dam is typically coated with $\sim$2 to 3 mm of syrup, when removed, effectively shortening the length of the fluid in the reservoir. However, in this geometry, propagation is insensitive to the length of the reservoir and this is below the spatial resolution of our forward model. Flow progression along the channel is recorded by time-lapse photographs from the side at a rate of 1 frame-per-second (FPS), and from above using a video camera (30 FPS). Experiments last between 15 and 180 s, with most 30 to 90 s in duration. We process the top-view videos to extract the flow front position over time. In the tracking algorithm, the video, rotated to align the flow direction left to right, is processed by manually identifying the starting position, scale, and center line in each video, and setting an appropriate threshold on the brightness of either the red, blue or green channel of the image to distinguish syrup from the background. In some cases, the time--distance profile of the flow front was smoothed to remove artifacts caused by the flash from the side-view camera. The flow front position is compared to a forward model of dam-break flow to invert for the rheological parameters. \par

\subsection{Extracting rheological parameters from experiments}
We follow the derivations of \citet{Liu1989} and \citet{Balmforth2007} for time evolution of flow thickness, $h$, of a Herschel--Bulkley fluid flowing down a slope:
\begin{subequations}
    \begin{align}
    \label{eq:flowPDE}
    \frac{\partial h}{\partial t} &= \frac{\partial}{\partial x} \left[ \left( \frac{\rho g}K \right)^{1/n} \frac{n \left| \sin{\theta} - \frac{\partial h}{\partial x} \cos{\theta} \right|^{1/n-1} Y^{1/n+1}}{\left( n+1 \right) \left( 2n+1 \right)} \left( \left( 2n+1 \right)h - nY \right) \left( \sin{\theta} - \frac{\partial h}{\partial x} \cos{\theta} \right) \right] \: , \\
    Y &= h - \frac{\tau_y}{\rho g | \sin{\theta} - \frac{\partial h}{\partial x} \cos{\theta} |} \: ,
    \end{align}
\end{subequations}
where $x$ is the along-channel distance, $\rho$ is the fluid density, $g$ is the acceleration due to gravity, and $\theta$ is the channel slope. \par 
This forward model is valid for non-inertial laminar flow of a gravitationally settling fluid experiencing no-slip conditions at the base of the flow and no-stress at the free surface. In the inertial regime, propagation of the flow front would be expected to scale with the shallow water speed $\sqrt{gH}$. Based on the short-time similarity solution of the flow front position given by \citet{Balmforth2007} for a Newtonian fluid, $X\sim0.2845 \hat{t}^{1/2}$, for $t = \frac L H \left( \frac{KL}{\rho g H^2} \right)^{1/n} \hat{t}$ and $x = L \hat{x}$, where $L$ is the reservoir length and $H$ the reservoir depth, we expect inertia would be dominant for $t \lesssim 10^{-5}-10^{-2}$ s, which is much shorter than the duration of our experiments. We confirm that all experiments are in the laminar regime by verifying that the Reynolds number, $\mathrm{Re}$, is below the cutoff for the onset of turbulence at $\mathrm{Re}\sim1000$. We calculate  $\mathrm{Re}$ for a free surface flow of a Herschel-Bulkley fluid in a rectangular channel as $\mathrm{Re} = \frac{\rho \bar{u}^{2-n}D^n}{(\tau_y/8) (D/\bar{u})^n + K ((3m+1)/(4m))^n 8^{n-1}}$, $m = \frac{n K (8 \bar{u}/D)^n}{\tau_y + K(8\bar{u}/D)^n}$ \citep{Madlener2009} where $\bar{u}$ is the mean velocity, $D$ is the hydrodynamic radius, $D = \frac{4H W}{2H+W}$, for channel width $W$ = 15 cm, and channel depth $H \sim h$. We find Reynolds numbers $\lesssim 10$, well below the cutoff. We confirm no-slip conditions via imaging through the base of the channel, which shows that bubbles are well coupled to the base (Supplementary Fig. \ref{fig:Bottom_view}). We neglect the effect of surface tension between the syrup and air, which becomes important for $h<2\sqrt{\frac{\Gamma}{\rho g}}\approx0.5$ cm \citep{deGennes2004}. Accordingly, we restrict our experiments to have a starting reservoir volume $H>4cm$. Finally, we require that the suspension be able to deform under its own weight, which requires that the Bingham number,  $\mathrm{B}<1$, where $\mathrm{B} \equiv \frac{\tau_y L}{\rho g H^2}$ \citep{Balmforth2007}. We prepared a suspension of $\phi_{solid}=0.46$, and find that for a flow height of 2.5 cm, it does not deform under it's own weight, suggesting that yield stress must exceed 42 Pa.\par

We solve equation \ref{eq:flowPDE} numerically using a finite-difference scheme on a centered 3-point stencil which is second-order accurate in space and we use two-step Runge-Kutte for second order accuracy in time. Initial conditions are fluid height $h = H$ where $x \leq L$ (within the reservoir), and $h = 0$ elsewhere. Boundary conditions are no-slip on the bottom, no-stress at the free surface, $\frac{\partial h}{\partial x}|_{x=x_L}=0$ (no inflow on low slope), and $h|_{x=x_R}=0$ where $x_R$ is chosen to be sufficiently far from the final position of the flow front such that flow does not reach the far boundary (domain length is 20\% longer than the final experimental flow length). The domain is discretized in space evenly with 52 grid points and with a constant time step of $\Delta t = 6.67\times 10^{-4}$ s. The forward model assumes that temperature, rheology, and density fluid of the fluid are constant in time and space. We allow slope to be non-zero in the model solution to account for sensitivity of flow advance rate to slope angles below our uncertainty for direct measurement ($\leq 0.5^{\circ}$). We extract the position of the flow front over time from the model by identifying where $h\approx0$, below a threshold of $1 \times 10^{-4}$ to account for numerical diffusion, smoothed linearly between grid points in $x$. \par 

We use the evolution of the flow front over time in the experiments as a constraining observation in our inversion for rheological parameters. We invert for rheological parameters ($K,\tau_y, n$) and the slope ($\theta$) using an Ensemble Kalman Filter (EnKF) approach \citep{Gregg2016, Zhan2017}. This is a probabilistic approach in which the forward model is instantiated many times with varying parameters that are modified iteratively to minimize the misfit between the model and data (typically $\sim$2000 data points). All samples are propagated together between iterations, which allows faster convergence than a random walk. We use Gaussian priors centered about a visually-determined close fit for the rheological parameters, and about zero for the slope. Standard deviations were 10\%, 10\%, 5\%, and 1$^{\circ}$ for the initial distributions of $K$, $\tau_y$, $n$, and $\theta$, respectively. EnKF simulations are initialized with 300 samples and run for 5 iterations; convergence is typically achieved within 4 iterations, insensitive to small changes in the initial parameters (Supplementary Fig. \ref{fig:EnKF}). In other Earth-science applications, EnKF is often used to update probabilities by assimilating more information through time, for example in volcano monitoring \citep{Gregg2016,Zhan2017}. We do not utilize this capability of EnKF; all information about the flow progression is included from the beginning. We use the Python implementation of the EnKF algorithm provided on GitHub by geoyanzhan3 \footnote{See https://github.com/geoyanzhan3/EnKF\textunderscore tutorial} \citep{Gregg2016,Zhan2017}. We report the numerically-determined maximum likelihood value, and uncertainty given by the 5\% and 95\% quantiles. We find uncertainty values of $\lesssim 20\%$ in $K$, $\lesssim 20 \%$ in $\tau_y$, and $\lesssim 10\%$ in $n$. \par

\section{Results}
We present results from 35 dam-break experiments spanning $0\leq\phi_{solid}\leq0.37$ and $0\leq\phi_{gas}\leq0.82$ (Fig. 2 and Table \ref{tbl:Experiments}). We find that, with increasing volume fraction of solids, relative consistency ($K_r = K/\mu$) increases, yield stress increases, and flow index decreases (more shear-thinning), consistent with previous studies \citep{Mader2013}. Increasing gas fraction in all cases increases relative consistency and yield stress. Gas-bearing experiments show both shear-thinning ($n<1$) and shear-thickening ($n>1$)  behavior, with increasing $n$ correlated with increasing $\mathrm{Ca}$. \par 

\begin{figure}
\begin{center}
\includegraphics[width=\textwidth]{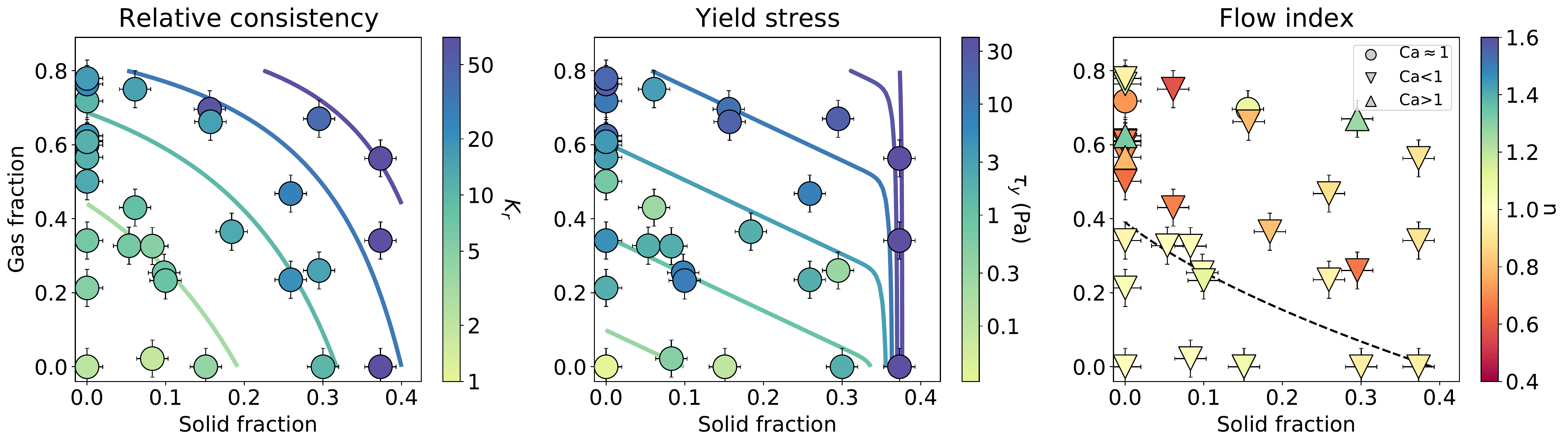}\\
\caption{Experimental rheology measurements plotted against solid (x-axis) and gas (y-axis) volume fractions, and the value of A) relative consistency, $K$, and B) yield stress, $\tau_y$ shown in color. Contours represent the best-fitting empirical models on the same color scale. In panel C), color represents the maximum capillary number, $\mathrm{Ca}$, and flow index, $n$, indicated by marker shape. The dashed black line indicates the critical volume fraction for the onset of shear-rate dependent behavior below which flow is shear-rate-independent, and above which shear-thinning behavior is expected at intermediate and low $\mathrm{Ca}$ and shear-thickening behavior is predicted for high $\mathrm{Ca}$. Data are reported in Table \ref{tbl:Experiments}.}
\label{fig:experiment_results}
\end{center}
\end{figure}

We use our measurements to parameterize the model framework presented in Section \ref{section:theoretical_framework}b, and find empirical expressions for $K_r$, $\tau_y$, and $n$ as functions of phase volume fractions and (where relevant) the maximum capillary number. We calculate maximum capillary number using equation \ref{eq:Ca} with the maximum observed bubble size from each experiment (0.5 to 8.2 mm) and estimate the maximum strain rate from experiment videos. Best fitting parameter values for these empirical models are determined using the L-BFGS-B algorithm for limited-memory, quasi-Newton, bound-constrained optimization \citep{Byrd1995} implemented in Python; uncertainty is estimated from the approximated Hessian matrix.

We find the following empirical relationships:
\begin{subequations}
    \begin{align}
    \label{eq:consistency}
    &K_r &&= \left( 1 - \frac{\phi_{solid}}{\phi_m} \right)^{-B_{solid}} \left( 1 - \phi_{gas} \right)^{-B_{gas}} \: , \\
    &\tau_y &&= 10^{C_1 (\phi_{solid} - \phi_{c,\tau_y})} + 10^{C_2 (\phi_{solid} + \phi_{gas} - \phi_{c, \tau_y})} \: , \label{eq:yieldstress} \\
    &n &&=
    \begin{cases}
      1, & \phi_{solid}(1-\phi_{gas}) + \phi_{gas} \leq \phi_{c,n} \\
      1 + (C_3 - C_4\mathrm{Ca})\left( \phi_{c,n} - \phi_{solid}(1-\phi_{gas}) - \phi_{gas} \right),  & \phi_{solid}(1-\phi_{gas}) + \phi_{gas}  > \phi_{c,n}
    \end{cases}  
    \label{eq:flowindex}
    \end{align}
\end{subequations}
where $\phi_m$=0.56$\pm$0.20, $B_{solid}$=2.74$\pm$1.56 and $B_{gas}$ = 1.98$\pm$0.09, $C_1$ = 80.0$\pm$10.9, $C_2$ = 1.98$\pm$0.23, the critical volume fraction for the onset of appreciable yield stress $\phi_{c,\tau_y}$ = 0.35$\pm$0.01, $C_3$ = 0.70$\pm$0.25, $C_4$=0.55$\pm$0.31, and the critical volume fraction for the onset of shear rate dependence $\phi_{c,n}$ = 0.39$\pm$0.12. Results from each experiment are plotted in Fig. 2 and compared to contours for the best-fitting model plotted on the same color scale. Misfit between the experiments and model are given in Table \ref{tbl:Experiments}.

\section{Discussion}
\subsection{Comparison with previous studies}
\subsubsection{Relative Consistency, $K_r$}
\begin{figure}
\begin{center}
\includegraphics[width=\textwidth]{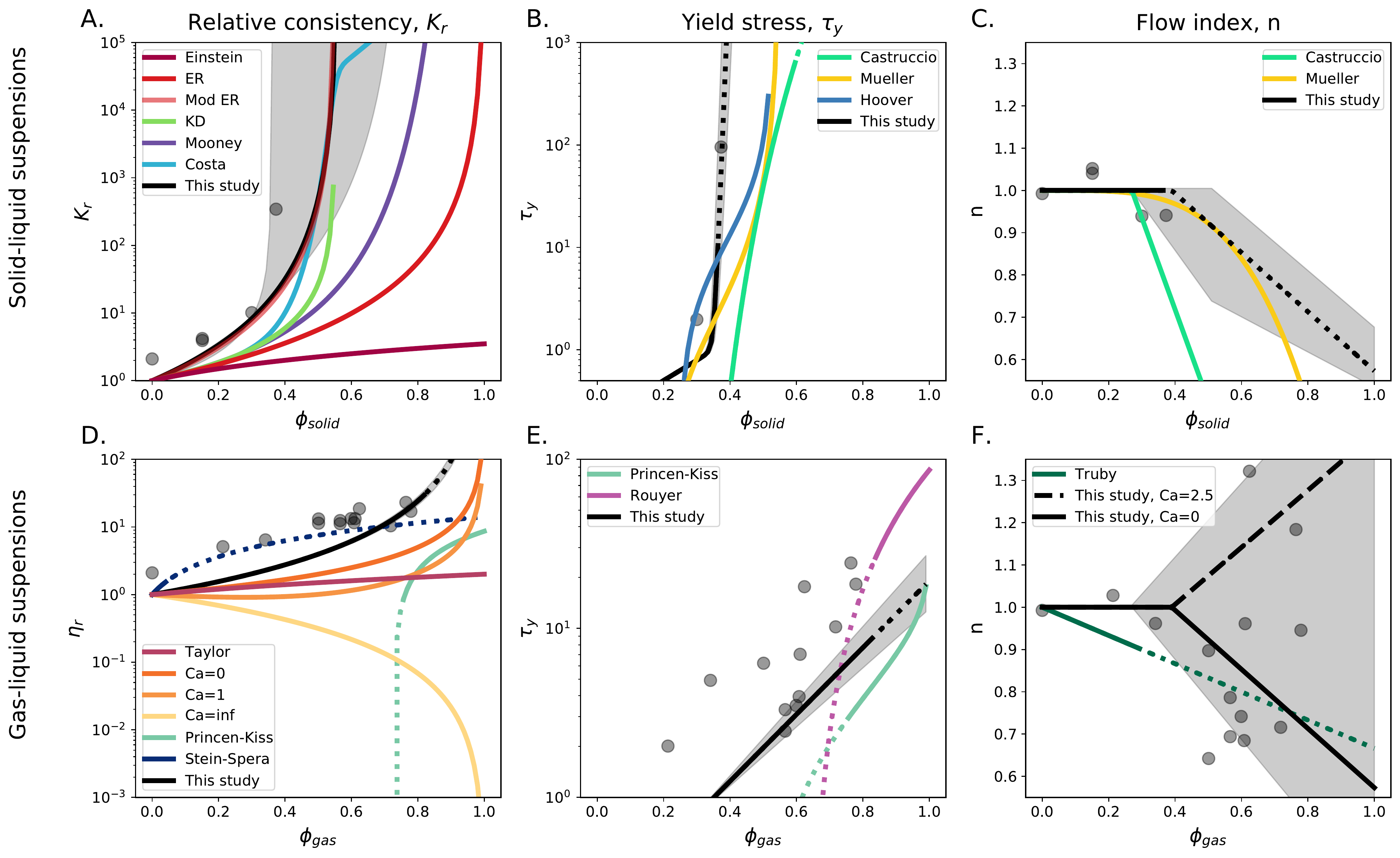}\\ 
\caption{Comparison between popular rheology models and the model proposed in this study. A and D) show relative consistency,  B and E) yield stress, and C and F) flow index in the two-phase end-member cases of A-C) solid-liquid only and D-F) gas-liquid only suspensions. Best-fit models for our experiments are plotted in black, with grey shaded regions highlighting model uncertainties. ER = \citet{Roscoe1952}, KD = \citet{Krieger1959}, Also see \citet{Mooney1951, Costa2009, Taylor1932, Llewellin2005, Princen1989, Stein1992, Castruccio2010, Mueller2010, Hoover2001, Rouyer2005, Truby2015} (See Table \ref{tbl:rheology_laws}). Dotted lines indicate extrapolations outside the calibration region for experimental studies. Grey markers are our two-phase experiments. Note that the model is constrained on the three-phase dataset so does not necessarily pass through this subset of the experiments.}
\label{fig:lit_rheo}
\end{center}
\end{figure}

Our expression for consistency (Eq. \ref{eq:consistency}) has the functional form presented in Eq. \ref{eq:theory_3phase_consistency}, such that it can be decomposed into separate functions for the two-phase components, facilitating comparison with previous studies (Fig. \ref{fig:lit_rheo}A and \ref{fig:lit_rheo}D, equations and parameters in Table \ref{tbl:rheology_laws}). For particle suspensions ($\phi_{gas}=0$), we find a $K_r(\phi_{solid})$ dependence that has the same functional form as that of \citet{Krieger1959} and \citet{Roscoe1952}. We do not have experiments for $\phi_{solid} > 0.37$ where relative consistency rapidly increases and a transition to another functional form, such as that in \citet{Costa2009}, may be appropriate. For bubble suspensions ($\phi_{solid}=0$) we find a $K_r(\phi_{gas})$ dependence that is stronger than most previous studies, with bubbles having a stronger stiffening effect on the suspensions. Furthermore, we find that in our experiments, bubbles always increase the relative consistency, regardless of capillary number, in contrast with previous studies \citep{Rust2002, Llewellin2002a}. \par 
We note that our consistency model (Eq. \ref{eq:consistency}) is optimized against the full three-phase dataset, hence three-phase interactions are likely to play a role in these discrepancies with models that are fitted against two-phase data. For instance, breakup of bubbles between interacting particles, and the formation of liquid-film bridges between particles may both amplify the effect of the bubble fraction on suspension rheology, and cause the rheology to be controlled by the smallest bubbles, which are in the low Ca regime, despite observations of deformation in the largest bubbles (Fig. \ref{fig:Bubble_def}), which imply $\mathrm{Ca}\gg 1$ locally.

\subsubsection{Yield Stress, $\tau_y$}
We select an empirical exponential-type relationship  for yield stress (Eq. \ref{eq:yieldstress}), which overlaps with previous two-phase studies (Fig. \ref{fig:lit_rheo}B and \ref{fig:lit_rheo}E). We observe a much stronger effect of particles compared to gas, but onset at around the same volume fraction. Yield stresses we observe in three-phase suspensions are higher than would be predicted by a linear superposition of the two-phase components, indicating again that interactions among the phases are important. At intermediate volume fractions we find moderately good agreement with \citet{Hoover2001} and \citet{Mueller2010}. The gas-liquid suspension experiments compare favorably to results by \citet{Princen1989} and \citet{Rouyer2005}, although there is limited overlap between regions of calibration.  

\subsubsection{Flow Index, n}
We fit flow index data using a piecewise continuous linear relationship after \citet{Castruccio2010} and \citet{Truby2015}. For particle suspensions, and for bubble suspensions at low $\mathrm{Ca}$, this formulation yields a shear-thinning behavior, which is consistent with previous relationships (Fig. \ref{fig:lit_rheo}C and \ref{fig:lit_rheo}F). With increasing $\mathrm{Ca}$ and gas fraction, it transitions to shear-thickening, a behavior reported in previous studies \citep{Pistone2012}.

\subsection{Data limitations}
We use the dam-break consistometer setup to measure the rheology of three-phase suspensions in order to avoid bubble breakdown which is common in conventional rheometers. This setup allows us to attain high bubble volume fractions, but requires numerical inversion and additional considerations for direct comparison to rheometry data collected via other methods. The behavior of Herschel-Bulkley fluids in confined channels and on unconfined slopes has been investigated theoretically and experimentally \citep{Coussot2014}, primarily using kaolin clay suspensions and Carbopol which show variable agreement with measurements collected in conventional rheometers. Most studies that use channel flow to investigate rheology use the approach to a final runout state \citep{Hogg2009}, the height profile \citep{Ancey2009, Huang1998, Vola2004}, the internal velocity \citep{Ancey2012, Andreini2012, Cantelli2009, Chambon2014}, or the runout distance with time \citep{Ancey2009, Balmforth2007, Cochard2009, Longo2016}. The internal velocity and height profiles observed in experiments using kaolin clay typically compare favorably to theory, except during the initial slumping stage, in which inertia is dominant, and near the flow front where significant curvature requires more sophisticated modeling \citep{Chambon2014, Huang1998}. In contrast, experiments using Carbopol show systematic offsets compared to conventional rheometers, possibly a result of basal slip, three-dimensional effects such as levee formation \citep{Ancey2012,Andreini2012}, or a scale effect introduced in the narrow gap of conventional rheometers compared to a relatively large microstructural length scale in Carbopol \citep{Chambon2014}. Our results use the runout distance with time to measure rheology which shows good agreement with conventional rheometry for experiments on corn syrup, xanthan gum, and kaolin suspensions \citep{Balmforth2007}. We expect our suspensions to behave similarly to two-phase kaolin particle suspensions, and further verify that basal slip and levee formation do not occur in our experiments. Given findings in previous work on two-phase suspensions in dam-break experiments and the good agreement between our experiments and existing particle-bearing suspensions, we do not expect systematic offsets in the rheometric parameters determined in our study compared to what would be measured in a conventional rheometer. \par

Our forward model is 1D (depth-integrated) and cannot capture any effects perpendicular to the channel direction. Visual inspection of the experiments suggests that there is not pronounced cross-channel flow. Our channel width, at 15 cm, is larger that that used by \citet{Balmforth2007} who find a 10 cm channel outperforms a 5 cm channel. We estimate the boundary layer width for a Herschel-Bulkley fluid, $\delta = \left(\frac{\left(1 + 1/n\right)^{n} K U^n H}{\tau_y}\right)^{n/(n+1)}$ \citep{Boujlel2012} at high Bingham number ($\mathrm{Bi}=\frac{\tau_y}{K} \left( \frac{U}{\delta} \right)^{-n}$), where $U$ is the maximum velocity along the channel, and scales with the height of the flow at low Bingham number. Our experiments fall in the intermediate Bingham number range, and we expect boundary layer thicknesses between $<$1 cm and 17.5 cm. Experiments with boundary layers larger than half the channel width (8.5 cm) are noted in Table \ref{tbl:Experiments}. The misfits between these experiments and the model do not show systematic differences nor are the misfits different in magnitude compared to experiments with low predicted boundary layer thicknesses. \par

The uncertainties on the values for $K_r$, $\tau_y$, and $n$ we derive from inversion of the experimental data are 25\%, 31\%, and 14\%, respectively, and the mismatch with models for those parameters (Eqs \ref{eq:consistency} -\ref{eq:flowindex}) are 35\%, 106\%, and 10\%. Our experiments show good repeatability at $\phi_{gas}\leq0.6$. If we compare experiments for which $\phi_{solid}$ and $\phi_{gas}$ are within $\pm$5\%, and de-trended $K_r$, $\tau_y$, and $n$ values using their respective models, we find ranges of 38\%, 149\%, and 20\% in those parameters, which are of the same order as uncertainties and mismatch. For $\phi_{gas}>0.6$, the data show greater variability with ranges of 68\%, 154\%, and 73\%, respectively. Data variability  may be related to experimental sources of uncertainty, such as small errors in flow density or height of fluid in the reservoir to which the forward model is very sensitive, or microstructural differences in the particle and bubble populations. Yield stress measurements at all bubble fractions show scatter between experiments and misfit to the model higher than the uncertainty predicted by EnKF model fitting, which may come from real variation among samples rather than trade-offs between fitting parameters.  \par
 
We see consistent deviations between the experimental data and the rheology model  in the range of $\phi_{gas}>0.6$, where the model consistently over-predicts both the relative consistency and flow index, perhaps due to trade-offs in the inversion of the experimental data. To test the importance of trade-offs in fitting parameters, we find the Pearson correlation coefficients between the EnKF instances after the final propagation (best fit for the fitting parameters) for each experiment. A value of 1 indicates a perfect positive correlation, -1 a perfect negative correlation, and 0 no correlation.The EnKF analyses reveal a moderate correlation between the slope angle and consistency, with an average correlation coefficient between EnKF samples for each experiment of C=0.63. We also find weak correlations between slope angle and yield stress (C=0.25) and slope angle and flow index (C=-0.44); a weak positive correlation between consistency and flow index in experiments with $\phi_{gas}>0.6$ (C=0.27); a moderate negative correlation when $\phi_{gas}<0.2$ (C=-0.40); and no apparent correlation at intermediate gas volume fractions $0.2<\phi_{gas}<0.6$ (C=-0.01). We do not observe correlations in the parameter inversions between yield stress and consistency (C=-0.06) or yield stress and flow index (C=0.11). For an example of the EnKF inversion results and match to experimental data see Supplementary Fig. \ref{fig:EnKF}. \par 

Our results consider strain rates up to $10^1$ 1/s and maximum $\mathrm{Ca}$ between $\approx10^{-2}$ and $10^{1}$. We find that bubbles lead to a higher effective viscosity than predicted by the Cross model \citep{Mader2013}, but our experiments do not cover a large-enough range of $\mathrm{Ca}$ to comment on asymptotic behavior that may occur at much higher or lower $\mathrm{Ca}$. Given the strong strain-rate dependence of the Herschel--Bulkley formulation, this model may not be appropriate for fluids experiencing strain rates far outside the calibration region ($\gg$10 1/s) as may occur in natural settings, including during conduit ascent in explosive eruptions or in fast-moving lava flows. \par

Additional uncertainty exists for the estimate of $\mathrm{Ca}$, which is difficult to define for a polydisperse bubble population whose exact bubble sizes and deformation cannot be directly observed. Estimates for the bubble size distribution are available in the supplementary material. Bubble deformation is observed for some of the largest bubbles experiencing the strongest shear, and thus experiencing the highest Ca conditions, imaged through the side walls of the channel (Supplementary Fig. \ref{fig:Bubble_def}). Such observations are limited to the exterior of the flow, which is necessarily dominated by edge effects. Better characterization of bubble sizes and deformations in the flow interior is an avenue for future work.

\subsection{Application to lava dynamics}
We demonstrate the implications of our new three-phase suspension rheology model with realistic parameters for lava flows through implementation to the channelized lava flow model PyFLOWGO \citep{Harris2001, Chevrel2018}. PyFLOWGO is a 1D lava flow advance model for the velocity and final runout distance of a cooling and crystallizing lava. The physical model uses a Bingham-plastic rheology of the form $\tau = \tau_y + \eta_{eff} \dot{\gamma}$. We add our experimentally constrained rheology model to several built-in parameterizations for the yield stress and effective viscosity. We calculate $\tau_y$ using Eq. 3b and the effective viscosity term $\eta_{eff} = K_r \mu \dot{\gamma}^{n-1}$ using Eqs. 3a and 3c. We compare our model with the commonly used two-phase modified Einstein-Roscoe relationship \citep{Roscoe1952}, and three-phase model of \citet{Phan-Thien1997}. \par 

We take the flow simulation parameters from the well-documented LSF1 lava flow from Mount Etna, Italy in 2001. We initiate the simulated lava with a low solid volume fraction of ${solid} = 0.1$. Solid fraction then increases, with further crystallization during emplacement, to $\phi_{solid} = 0.15$ when using our rheological relations, and $\phi_{solid} = 0.3$ for the modified Einstein-Roscoe and \citet{Phan-Thien1997} relationships. Gas fraction begins and remains at $\phi_{gas}=0.65$. The simulated flow follows the path of steepest descent in the pre-eruptive topography. The initial flux is 20 m$^3$/s. We use a temperature-dependent melt viscosity using the VFT equation (a = -4.827, b = 5997, c = -330.3) \citep{Cordonnier2016, Coltelli2007, Lombardo2006}. We tune the model parameters of initial channel width (set to 40 m) and temperature (set to 1223.15 K) to obtain the observed flow height (10 m) and mean velocity ($10^{-3}$ to $4\times10^{-2}$ m/s) when using our new rheology model. The width and initial temperature are held constant across all simulations. To predict the effective viscosity, we use a reference strain rate of $10^{-3}$ s$^{-1}$ for this lava flow, estimated from observed velocities and flow thickness. We approximate the shear rate dependence using the capillary number calculated for a 1 cm radius bubble and a surface tension of air in basalt of 0.37 N/m \citep{Walker1981}. As the flow cools, the melt viscosity,  crystal content, and effective bulk viscosity are updated every step. \par

\begin{figure}
\begin{center}
\includegraphics[width=2.4in]{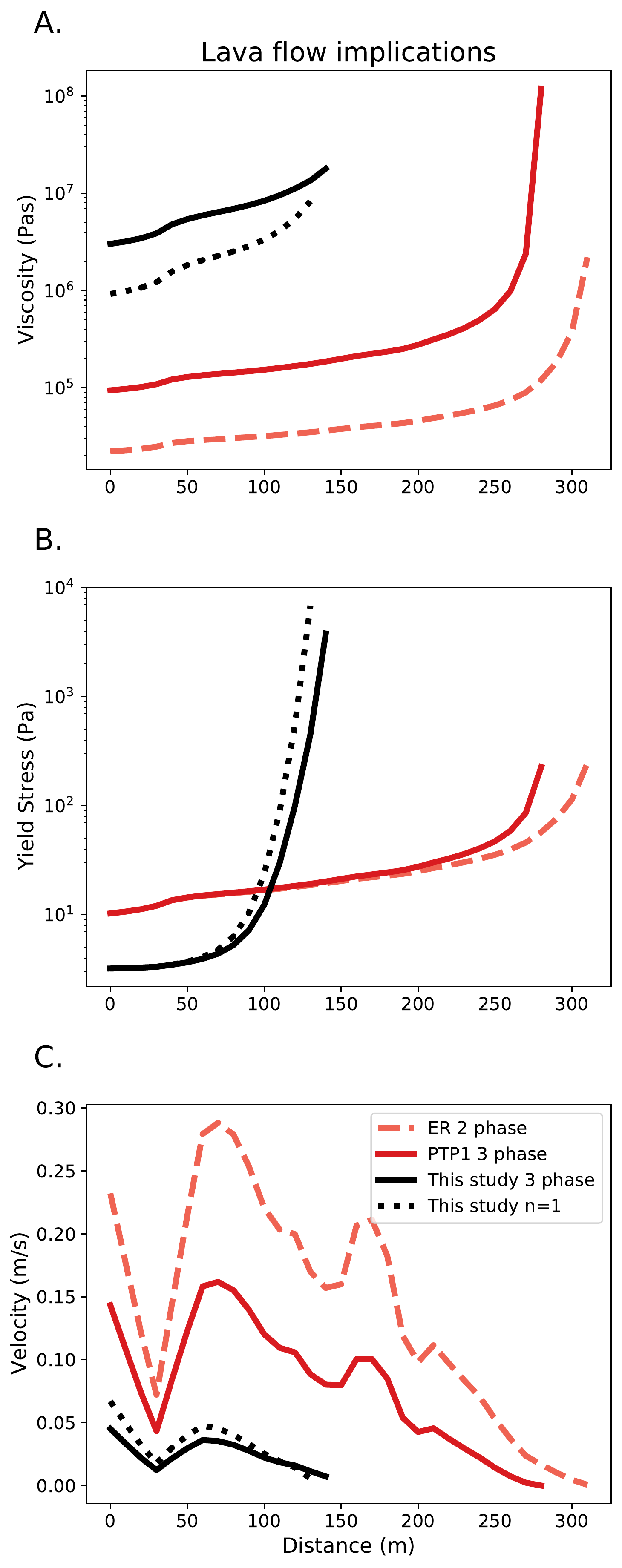}\\
\caption{Implications of the experimentally determined three-phase rheology on predictions of lava flows. Our model results in a higher relative effective viscosity (A), a sharper increase in yield stress as particle-bubble interactions become important (B), and lower mean velocity (C) compared to the models of \citet{Roscoe1952} (two-phase) or \citet{Phan-Thien1997} (three-phase). We also highlight the contrast between assuming n=1 (Bingham-plastic rheology) and the Herschel-Bulkley rheology (n$\approx$0.8) that results in a decrease in effective viscosity by a factor of 2 at low crystal fractions.}
\label{fig:PyFLOWGO_Confort15}
\end{center}
\end{figure}

Our model predicts a higher effective viscosity compared to that of \citet{Phan-Thien1997} or \citet{Roscoe1952} by more than an order of magnitude, and yields a mean velocity that is $\sim$3 times slower (Fig. \ref{fig:PyFLOWGO_Confort15}). Additionally, our yield strength model diverges from that based on \citet{Dragoni1989} $\sim$80 m down flow, when the crystal fraction increases enough to support significant bubble-crystal interactions. Although shear-thinning behavior is not included in PyFLOWGO, we compare effective viscosities predicted by our model assuming a Bingham-plastic rheology ($n=1$) and with a dynamic value for $n$ in response to changing crystal fraction (while maintaining a constant shear rate). We highlight that for the large bubble volume fraction of some natural lavas, $n$ deviates from one (initially $n\approx0.82$) and can result in a factor of two difference in effective viscosity, and a 15\% difference in runout distance . This effect may be more pronounced in modeling that includes a physical model for shear-rate dependence in the flow.\par 

\section{Conclusion}
We use a dam-break setup to measure the rheology of three-phase suspensions, scaled to apply to magma and lava. Through the use of a chemical reaction between baking soda and citric acid, we are able to create corn syrup suspensions with bubble volume fractions up to 0.82 and particle volume fractions up to 0.37. Strain rates in our experiments range up to $10^1$ 1/s and maximum $\mathrm{Ca}$ between $\approx10^{-2}$ and $10^{1}$, relevant to basaltic lava flows. We invert for a Herschel--Bulkley rheology using a depth-integrated 1D forward model \citep{Liu1989} and the probabilistic Ensemble Kalman Filter approach. \par 
Based on our results, we develop a suite of models for the Herschel--Bulkeley parameters, calibrated across the whole dataset, which are presented in equations \ref{eq:consistency}-\ref{eq:flowindex}. Combined with equation \ref{eq:herschel-bulkley}, this constitutes a rheological model for three phase suspensions. Our results show good agreement with existing literature for particle-bearing suspensions with a strongly non-linear increase in viscosity with increasing particle fraction, development of yield stresses at particle fractions $\phi_{solid}\gtrsim0.35$, and development of shear-thinning behavior at particle fractions $\phi_{solid}\gtrsim0.39$. Our findings indicate that suspended gas leads to a non-linear increase in viscosity with increasing bubble fraction, larger than predicted by previous work, and development of yield stresses and shear rate-dependent behavior at bubble fractions, or combined bubble and particle fractions, similar to those observed for particle-only suspensions. We find that shear-rate dependence of bubble-bearing two- and three-phase suspensions are shear thinning in the low capillary number regime and become shear thickening with increasing capillary number. This study contributes an explicit incorporation of three-phase interactions into a novel model for yield stress, and a parameterization for the inclusion of capillary number in flow index. \par 
We highlight the implications of our model for magma/lava dynamics through application to the lava flow emplacement model PyFLOWGO, and show higher viscosities and lower predicted velocities for surface lava flows as a result of the stiffening behavior of concentrated bubble suspensions at low and intermediate capillary numbers. Our findings underscore the need to incorporate three-phase rheology into simulations of magma or lava flow used to address fundamental science questions as well as hazard assessment and mitigation.\vskip6pt

\enlargethispage{20pt}

\dataccess{Experiment videos available at https://doi.org/10.5281/zenodo.4685257. data processing code available at https://doi.org/10.5281/zenodo.4707969 or https://github.com/JanineBirnbaum18/3-phase-flow.}

\aucontribute{J.B. carried out the experiments and data analysis. E.L. conceived of and supervised the experiments and data analysis. E.W.L. contributed to the interpretation of the results. J.B. took the lead in writing the manuscript. All authors provided critical feedback and helped shape the research, analysis and manuscript.}

\competing{The authors have no competing interests to declare.}

\funding{This work was supported by NSF under awards NSF EAR-1654588 and NSF EAR-1929008, and by NERC under award NE/T009594/1.}

\ack{We thank J. Hammer and A. Whittington for their constructive comments, and Lamont-Doherty's Secondary School Field Research Program participants R. Burgos, M. Diakite, and L. Lyons for their assistance in conducting experiments. We thank two anonymous reviewers for their constructive suggestions which helped improve and clarify the manuscript.}

\bibliography{references}

\newpage
\section*{Supplementary Material}
\renewcommand{\thefigure}{S\arabic{figure}}
\setcounter{figure}{0}

\section{Supplementary Material}
\subsection{Supplementary figure descriptions}
We characterize the bubble size distributions of our experiments using optical imaging at two different spatial scales. First, we consider the smallest size fraction of bubbles in undeformed syrup suspensions using optical microscopy which allows us to resolve bubbles between 20 $\mu$m and 2 mm in diameter. Second, we characterize the size distribution and shear-induced elongation of bubbles $\gtrsim$ 1.5 mm that are visible without magnification through the walls of the clear channel. \par 

The smallest bubble size fractions are determined from photomicrographs analyzed using a MATLAB circle-finding algorithm and FOAMS (Shea et al. 2010). Results show a power-law distribution of bubble number density, shown in Fig. S1 for an undeformed sample with only gas suspended in liquid (no solids). In the presence of particles, heterogeneous nucleation is observed with small bubbles ($\lesssim 50 \mu$m in diameter) nucleating and growing on the uneven surfaces of the particles. This population shows greater spatial heterogeneity, obvious in panel B of Fig. S1 in which a region with fewer particles shows fewer and larger bubbles, while an adjacent area shows a greater density of small bubbles. The high density of bubbles on the particle surfaces introduce significant errors to the size distribution estimate for three-phase suspensions, making quantitative comparison challenging. We might additionally expect the population size to vary in response to shear, but deformation and coalescence at this size fraction are hindered by the relatively high surface tension. \par 

To characterize the larger bubbles visible through the side walls of the channel, we again utilize FOAMS (Shea et al. 2010) to calculate bubble volume, aspect ratio, and elongation direction distributions. We compare the bubbles found in the undeformed reservoir with the flow front after 8 min to show a change in the elongation direction of large bubbles (Fig. S2). The reservoir and the flow front show similar size distributions with a logarithmic decrease in bubble number density with increasing size at a comparable rate to the smaller size fraction (measured on two different sample suspensions). Both populations show slight elongations (mean aspect ratio $\approx$ 0.6), but the bubbles in the reservoir show a preferential vertical orientation as bubbles rise due to buoyancy, while the bubbles at the flow front are aligned with the flow direction, an approximately 90$^\circ$ rotation. \par

We also add a photograph in Fig. S3 taken from the underside of the channel looking up at the flow. It shows bubbles in contact with the bottom of the channel, without the formation of a segregated bubble-free layer. \par

Fig. S4 shows the relative error between experiments and the empirical model, normalized by the experimental results. We also include an example experiment inversion including the posterior distributions of $K_r$,$\tau_y$,$n$,and $\theta$ which show strong covariance between the slope and relative consistency, a good fit between observations and the maximum likelihood model, and convergence of the EnKF inversion after 3 iterations. Finally, we report each relationship shown in Fig. 3.
Fig S5 shows a typical example of inversion results for a flow including A.) the posterior distributions and covariations between rheological parameters ($K_r$,$\tau_y$,$n$) and slope ($\theta$); B.) the flow front propagation of the experiment (red) and model (blue) with confidence bounds from forward models using the 5th and 95th quantiles of the posterior distributions; and C.) the convergence of the EnKF inversion. 
Supplementary references
T. Shea, B. F. Houghton, L. Gurioli, K. V. Cashman, J. E. Hammer, B. J. Hobden (2010). Textural studies of vesicles in volcanic rocks: An integrated methodology. J. Volcanol. Geotherm. Res. 190, 271–289.

\renewcommand{\thefigure}{S\arabic{figure}}
\setcounter{figure}{0}
\renewcommand{\thetable}{S\arabic{table}}
\setcounter{table}{1}

\begin{figure}[!htb]
\begin{center}
\includegraphics[width=6in]{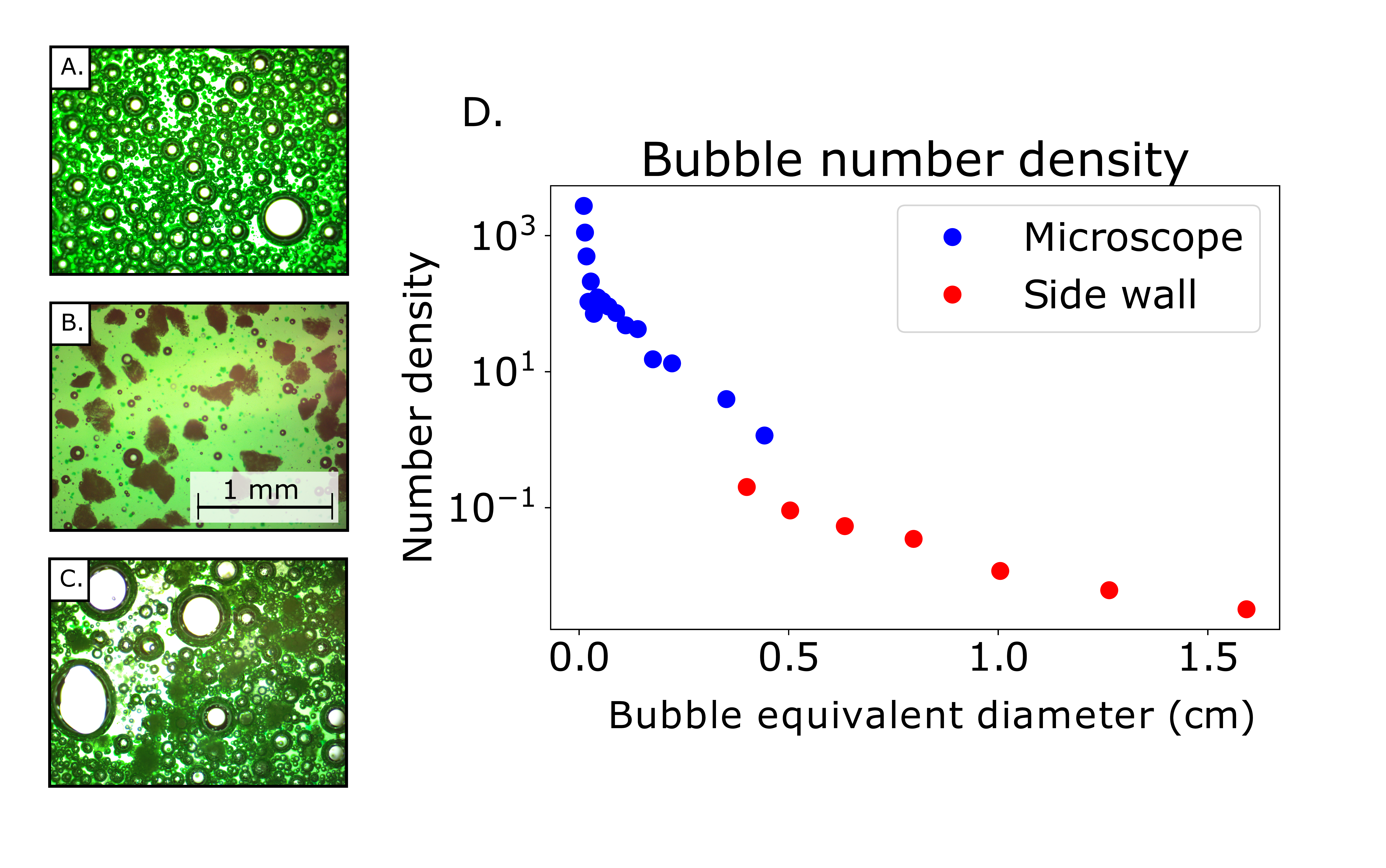}\\
\caption{Photomicrographs of A) bubbles, B) particles, and C) both bubbles and particles. Note the change in size distribution between the left and right sides of C) as a result of small-scale heterogeneities in particle distribution. D) Summarizes the bubble number density of representative solutions where small bubbles were characterized under the microscope and large bubbles were identified optically through the clear sides of the channel.}
\label{fig:BND}
\end{center}
\end{figure}

\begin{figure} 
\begin{center}
\includegraphics[width=\textwidth]{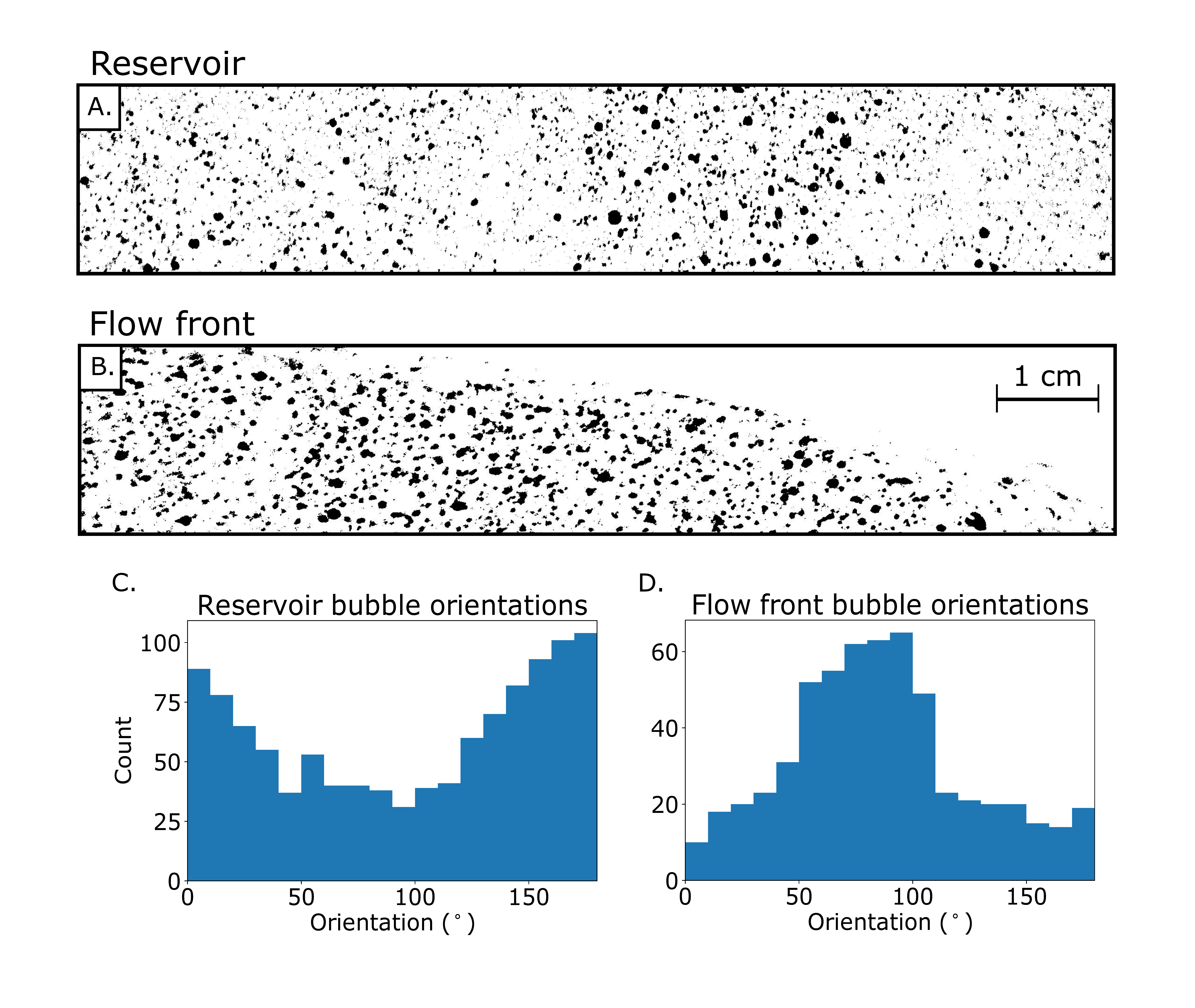}\\
\caption{Binary images of bubbles (black) suspended in syrup (white) as seen through the clear channel walls in the A) undeformed fluid reservoir, and B) the flow front of the same flow. A) and B) share the same scale. C) and D) show the respective bubble elongation orientations, highlighting a 90$^\circ$ rotation from vertically-deformed bubbles (C) to bubbles aligned with horizontal flow to the right (D).}
\label{fig:Bubble_def}
\end{center}
\end{figure}

\begin{figure} 
\begin{center}
\includegraphics[width=\textwidth]{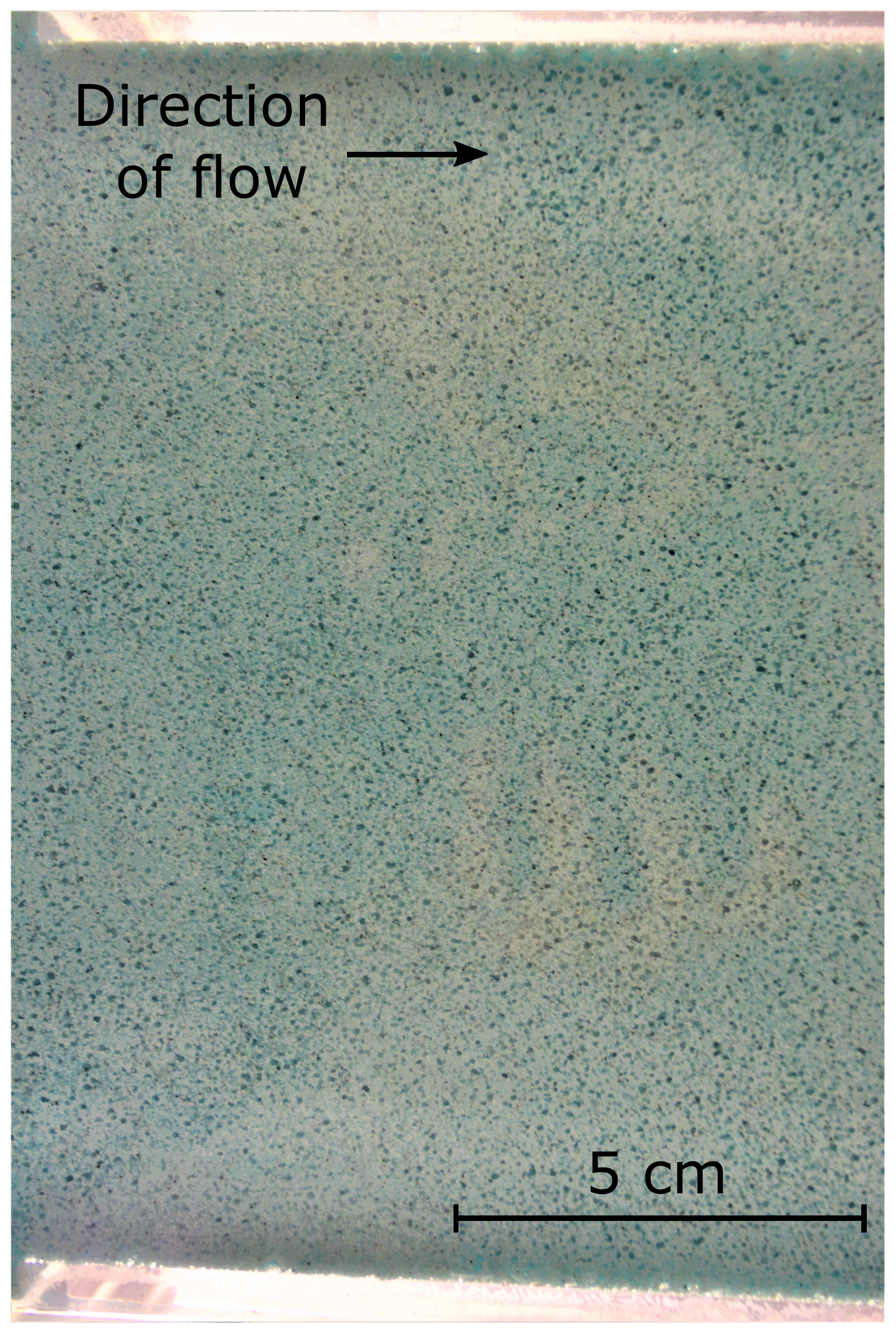}\\
\caption{View of an example flow from below showing bubbles in contact with the bottom of the channel. Bubbles do not show preferred orientation in the direction of flow, consistent with a no-slip condition at the base of the flow.}
\label{fig:Bottom_view}
\end{center}
\end{figure}

\begin{figure}
\begin{center}
\includegraphics[width=\textwidth]{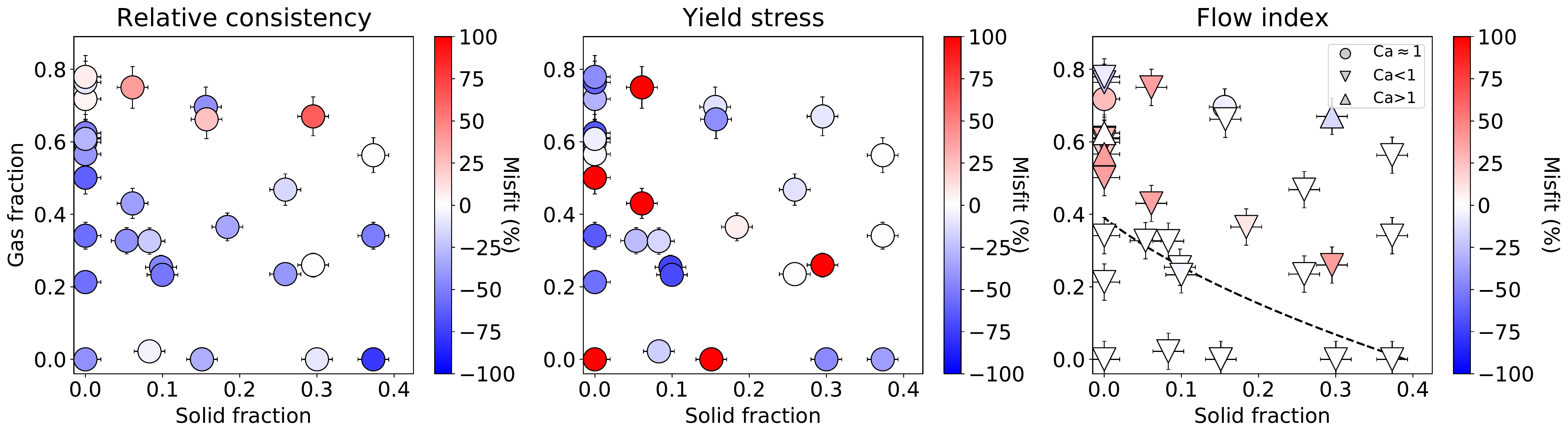}\\
\caption{Relative misfit between experimental measurements for rheological parameters and best-fit model for A) relative consistency, $K$, B) yield stress, $\tau_y$, and C) flow index, $n$.}
\label{fig:Error}
\end{center}
\end{figure}
 
\begin{figure}
\begin{center}
\includegraphics[width=\textwidth]{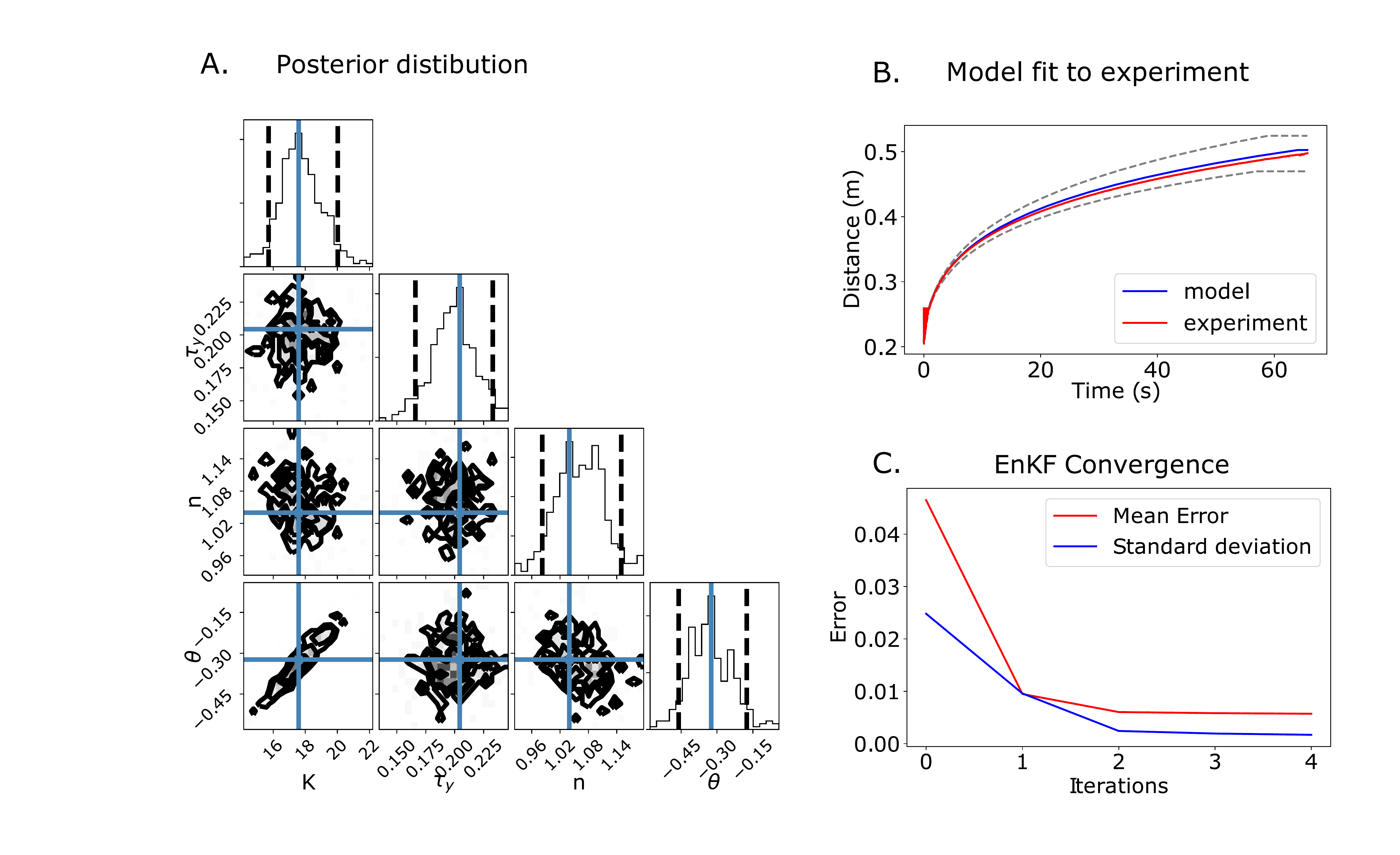}\\
\caption{Representative Ensemble Kalman Filter (EnKF) inversion. A) shows a corner plot of the posterior distribution that highlight the covariation between pairs of parameters, most clearly the relationship between relative consistency and slope. B) forward model of best fit parameters (blue) and experimental data (red), with confidence intervals (dashed grey). C) EnKF convergence with 300 ensembles over five iterations.}
\label{fig:EnKF}
\end{center}
\end{figure}

\clearpage
\begin{table}
\rotatebox{90}{\resizebox{545pt}{!}{
\begin{tabular}{l|l|l|l|l|l|l|l|l|l|l|l|l|l|l|l|l}
Video & Temp & Height & Syrup visc & Density & Solid & Gas & Max bubble & Max strain & Capillary & Relative & Error & Yield stress & Error & Flow & Error & Boundary-layer \\ 
& ($^{\circ}$C) & (m) & (Pas) & (kg/m$^3$) & & & radius (m) & rate (1/s) & number & consistency & (\%) & (Pa) & (\%) & index & (\%) & thickness (m) \\
\hline
MVI\_0006 & 28.8 & 0.095 & 4.29 & 535 & 0 & 0.61 & 0.006 & 1.16 & 0.40 & 13.27 $\pm \begin{tabular}{@{}c@{}} 1.29 \\ 1.54 \end{tabular}$ & -39.37 & 7.03 $\pm \begin{tabular}{@{}c@{}} 1.28 \\ 1.03 \end{tabular}$ & -39.03 & 0.96 $\pm \begin{tabular}{@{}c@{}} 0.05 \\ 0.10 \end{tabular}$ & 0 & 0.057 \\
MVI\_0007 & 22.3 & 0.042 & 8.00 & 874 & 0.30 & 0.67 & 0.007 & 5.32 & 3.46 & 47.76 $\pm \begin{tabular}{@{}c@{}} 7.97 \\ 6.96 \end{tabular}$ & 31.39 & 16.96 $\pm \begin{tabular}{@{}c@{}} 2.07 \\ 2.97 \end{tabular}$ & 0 & 1.14 $\pm \begin{tabular}{@{}c@{}} 0.10 \\ 0.06 \end{tabular}$ & 18.95 & 0.006 \\
MVI\_0010 & 23.4 & 0.100 & 7.12 & 458 & 0.30 & 0.67 & 0.007 & 1.83 & 1.20 & 40.61 $\pm \begin{tabular}{@{}c@{}} 4.33 \\ 6.58 \end{tabular}$ & 63.49 & 22.48 $\pm \begin{tabular}{@{}c@{}} 2.28 \\ 4.01 \end{tabular}$ & -9.64 & 1.27 $\pm \begin{tabular}{@{}c@{}} 0.07 \\ 0.10 \end{tabular}$ & -14.21 & 0.009 \\
MVI\_0011 & 23.9 & 0.041 & 6.76 & 1028 & 0.30 & 0.26 & 0.002 & 3.11 & 0.58 & 14.76 $\pm \begin{tabular}{@{}c@{}} 1.25 \\ 2.35 \end{tabular}$ & 0 & 0.31 $\pm \begin{tabular}{@{}c@{}} 0.05 \\ 0.05 \end{tabular}$ & 709.93 & 0.66 $\pm \begin{tabular}{@{}c@{}} 0.05 \\ 0.05 \end{tabular}$ & 38.90 & 0.041 \\
MVI\_0012 & 24.2 & 0.110 & 6.56 & 351 & 0.06 & 0.75 & 0.003 & 2.05 & 0.44 & 14.98 $\pm \begin{tabular}{@{}c@{}} 0.69 \\ 2.05 \end{tabular}$ & 38.66 & 3.30 $\pm \begin{tabular}{@{}c@{}} 0.35 \\ 0.58 \end{tabular}$ & 134.46 & 0.57 $\pm \begin{tabular}{@{}c@{}} 0.05 \\ 0.04 \end{tabular}$ & 37.01 & 0.110* \\
MVI\_0013 & 24.4 & 0.055 & 6.43 & 794 & 0.06 & 0.43 & 0.002 & 4.92 & 0.79 & 8.28 $\pm \begin{tabular}{@{}c@{}} 0.98 \\ 0.95 \end{tabular}$ & -37.95 & 0.27 $\pm \begin{tabular}{@{}c@{}} 0.05 \\ 0.04 \end{tabular}$ & 569.50 & 0.68 $\pm \begin{tabular}{@{}c@{}} 0.05 \\ 0.06 \end{tabular}$ & 36.24 & 0.055 \\
MVI\_0703 & 28.4 & 0.060 & 4.44 & 696 & 0 & 0.5 & 0.005 & 1.26 & 0.33 & 11.36 $\pm \begin{tabular}{@{}c@{}} 0.93 \\ 1.96 \end{tabular}$ & -47.81 & 6.22 $\pm \begin{tabular}{@{}c@{}} 1.07 \\ 0.88 \end{tabular}$ & -54.17 & 0.90 $\pm \begin{tabular}{@{}c@{}} 0.07 \\ 0.07 \end{tabular}$ & 0 & 0.071 \\
MVI\_0707 & 30.1 & 0.052 & 3.85 & 697 & 0 & 0.5 & 0.003 & 1.50 & 0.18 & 13.13 $\pm \begin{tabular}{@{}c@{}} 1.83 \\ 0.98 \end{tabular}$ & -62.32 & 0.70 $\pm \begin{tabular}{@{}c@{}} 0.13 \\ 0.11 \end{tabular}$ & 162.54 & 0.64 $\pm \begin{tabular}{@{}c@{}} 0.05 \\ 0.05 \end{tabular}$ & 37.93 & 0.052 \\
MVI\_0708 & 30.5 & 0.051 & 3.73 & 605 & 0 & 0.57 & 0.004 & 1.50 & 0.27 & 12.43 $\pm \begin{tabular}{@{}c@{}} 1.84 \\ 1.15 \end{tabular}$ & -48.60 & 2.47 $\pm \begin{tabular}{@{}c@{}} 0.35 \\ 0.42 \end{tabular}$ & 0 & 0.69 $\pm \begin{tabular}{@{}c@{}} 0.06 \\ 0.05 \end{tabular}$ & 22.58 & 0.051 \\
MVI\_0709 & 30.4 & 0.051 & 3.76 & 559 & 0 & 0.6 & 0.004 & 1.86 & 0.34 & 13.29 $\pm \begin{tabular}{@{}c@{}} 1.21 \\ 1.88 \end{tabular}$ & -39.76 & 3.50 $\pm \begin{tabular}{@{}c@{}} 0.34 \\ 0.69 \end{tabular}$ & 0 & 0.74 $\pm \begin{tabular}{@{}c@{}} 0.05 \\ 0.07 \end{tabular}$ & 13.83 & 0.175* \\
MVI\_0710 & 18.8 & 0.070 & 12.14 & 517 & 0 & 0.62 & 0.008 & 2.73 & 3.40 & 18.62 $\pm \begin{tabular}{@{}c@{}} 2.31 \\ 2.58 \end{tabular}$ & -48.77 & 17.64 $\pm \begin{tabular}{@{}c@{}} 2.01 \\ 2.61 \end{tabular}$ & -65.61 & 1.32 $\pm \begin{tabular}{@{}c@{}} 0.08 \\ 0.11 \end{tabular}$ & 0 & 0.070 \\
MVI\_0711 & 18.8 & 0.046 & 12.14 & 597 & 0 & 0.57 & 0.006 & 2.87 & 2.75 & 11.17 $\pm \begin{tabular}{@{}c@{}} 1.75 \\ 1.32 \end{tabular}$ & -41.36 & 3.30 $\pm \begin{tabular}{@{}c@{}} 0.51 \\ 0.55 \end{tabular}$ & -2.88 & 0.79 $\pm \begin{tabular}{@{}c@{}} 0.05 \\ 0.06 \end{tabular}$ & 38.54 & 0.046 \\
MVI\_0712 & 18.8 & 0.055 & 12.14 & 387 & 0 & 0.72 & 0.002 & 3.59 & 1.09 & 10.40 $\pm \begin{tabular}{@{}c@{}} 1.46 \\ 1.69 \end{tabular}$ & 4.27 & 10.21 $\pm \begin{tabular}{@{}c@{}} 1.64 \\ 1.80 \end{tabular}$ & -30.43 & 0.72 $\pm \begin{tabular}{@{}c@{}} 0.06 \\ 0.05 \end{tabular}$ & 26.56 & 0.105* \\
MVI\_0713 & 20.4 & 0.134 & 9.95 & 324 & 0 & 0.76 & 0.002 & 1.33 & 0.30 & 23.05 $\pm \begin{tabular}{@{}c@{}} 2.63 \\ 3.21 \end{tabular}$ & -10.06 & 24.37 $\pm \begin{tabular}{@{}c@{}} 3.26 \\ 2.98 \end{tabular}$ & -60.99 & 1.18 $\pm \begin{tabular}{@{}c@{}} 0.06 \\ 0.09 \end{tabular}$ & -24.31 & 0.017 \\
MVI\_0717 & 19.5 & 0.115 & 11.11 & 303 & 0 & 0.78 & 0.002 & 0.99 & 0.30 & 17.04 $\pm \begin{tabular}{@{}c@{}} 1.63 \\ 2.60 \end{tabular}$ & 7.52 & 18.26 $\pm \begin{tabular}{@{}c@{}} 1.67 \\ 3.01 \end{tabular}$ & -45.21 & 0.95 $\pm \begin{tabular}{@{}c@{}} 0.04 \\ 0.09 \end{tabular}$ & -7.08 & 0.064 \\
MVI\_0718 & 19.5 & 0.054 & 11.11 & 539 & 0 & 0.61 & 0.002 & 2.55 & 0.57 & 11.55 $\pm \begin{tabular}{@{}c@{}} 1.45 \\ 1.78 \end{tabular}$ & -29.11 & 3.95 $\pm \begin{tabular}{@{}c@{}} 0.74 \\ 0.47 \end{tabular}$ & -6.70 & 0.68 $\pm \begin{tabular}{@{}c@{}} 0.05 \\ 0.05 \end{tabular}$ & 27.11 & 0.054 \\
MVI\_0726 & 18.1 & 0.051 & 13.32 & 417 & 0.16 & 0.7 & 0.002 & 3.21 & 0.96 & 62.83 $\pm \begin{tabular}{@{}c@{}} 11.09 \\ 9.38 \end{tabular}$ & -43.63 & 13.17 $\pm \begin{tabular}{@{}c@{}} 2.26 \\ 1.78 \end{tabular}$ & -12.49 & 1.08 $\pm \begin{tabular}{@{}c@{}} 0.06 \\ 0.07 \end{tabular}$ & -6.67 & 0.009 \\
MVI\_0739 & 22.5 & 0.094 & 7.83 & 938 & 0.05 & 0.33 & 0.002 & 3.45 & 0.51 & 6.38 $\pm \begin{tabular}{@{}c@{}} 0.68 \\ 0.77 \end{tabular}$ & -42.56 & 2.00 $\pm \begin{tabular}{@{}c@{}} 0.34 \\ 0.33 \end{tabular}$ & -26.50 & 1.00 $\pm \begin{tabular}{@{}c@{}} 0.06 \\ 0.07 \end{tabular}$ & 0 & 0.085* \\
MVI\_0740 & 24.1 & 0.130 & 6.62 & 470 & 0.16 & 0.66 & 0.002 & 1.20 & 0.20 & 15.66 $\pm \begin{tabular}{@{}c@{}} 1.89 \\ 1.43 \end{tabular}$ & 23.58 & 20.96 $\pm \begin{tabular}{@{}c@{}} 2.17 \\ 3.35 \end{tabular}$ & -43.97 & 0.79 $\pm \begin{tabular}{@{}c@{}} 0.06 \\ 0.07 \end{tabular}$ & 0 & 0.120* \\
MVI\_0741 & 23.8 & 0.091 & 6.83 & 1386 & 0.37 & 0 & 0 & 2.82 & 0.00 & 343.28 $\pm \begin{tabular}{@{}c@{}} 51.85 \\ 54.69 \end{tabular}$ & -78.08 & 95.43 $\pm \begin{tabular}{@{}c@{}} 15.62 \\ 14.63 \end{tabular}$ & -28.53 & 0.94 $\pm \begin{tabular}{@{}c@{}} 0.05 \\ 0.07 \end{tabular}$ & 1.59 & 0.052 \\
MVI\_0742 & 23.0 & 0.120 & 7.42 & 605 & 0.37 & 0.56 & 0.002 & 2.78 & 0.52 & 101.98 $\pm \begin{tabular}{@{}c@{}} 13.82 \\ 15.45 \end{tabular}$ & 0 & 44.33 $\pm \begin{tabular}{@{}c@{}} 6.57 \\ 4.82 \end{tabular}$ & 35.77 & 0.90 $\pm \begin{tabular}{@{}c@{}} 0.09 \\ 0.05 \end{tabular}$ & 0 & 0.096* \\
MVI\_0743 & 22.6 & 0.095 & 7.75 & 913 & 0.37 & 0.34 & 0.001 & 2.72 & 0.13 & 148.46 $\pm \begin{tabular}{@{}c@{}} 14.14 \\ 25.25 \end{tabular}$ & -51.32 & 38.56 $\pm \begin{tabular}{@{}c@{}} 7.52 \\ 4.26 \end{tabular}$ & 30.07 & 0.90 $\pm \begin{tabular}{@{}c@{}} 0.04 \\ 0.07 \end{tabular}$ & 0 & 0.093* \\
MVI\_0744 & 23.4 & 0.108 & 7.12 & 883 & 0.18 & 0.37 & 0.001 & 1.73 & 0.15 & 13.20 $\pm \begin{tabular}{@{}c@{}} 1.49 \\ 1.23 \end{tabular}$ & -34.89 & 2.02 $\pm \begin{tabular}{@{}c@{}} 0.30 \\ 0.33 \end{tabular}$ & 6.70 & 0.80 $\pm \begin{tabular}{@{}c@{}} 0.07 \\ 0.05 \end{tabular}$ & 9.73 & 0.108* \\
\end{tabular}
}}
\caption{Table of experimental data, inverted rheological parameters, and misfit from the three-phase rheology model. Errors of 0 fit the model within the 5 and 95$\%$ confidence interval. Boundary layer thicknesses marked with a * are larger than the channel half-width (7.5 cm), and may be expected to have significant edge effects. }
\label{tbl:Experiments}
\end{table}

\clearpage
\begin{table}
\rotatebox{90}{\resizebox{600pt}{!}{
\begin{tabular}{l|l|l|l|l|l|l|l|l|l|l|l|l|l|l|l|l}
Video & Temp & Height & Syrup visc & Density & Solid & Gas & Max bubble & Max strain & Capillary & Relative & Error & Yield stress & Error & Flow & Error & Boundary-layer \\ 
& ($^{\circ}$C) & (m) & (Pas) & (kg/m$^3$) & & & radius (m) & rate (1/s) & number & consistency & (\%) & (Pa) & (\%) & index & (\%) & thickness (m) \\
\hline
MVI\_0745 & 24.0 & 0.101 & 6.69 & 1063 & 0.26 & 0.24 & 0.000 & 0.90 & 0.02 & 19.84 $\pm \begin{tabular}{@{}c@{}} 1.65 \\ 2.38 \end{tabular}$ & -40.67 & 2.07 $\pm \begin{tabular}{@{}c@{}} 0.25 \\ 0.44 \end{tabular}$ & 0 & 0.95 $\pm \begin{tabular}{@{}c@{}} 0.07 \\ 0.06 \end{tabular}$ & 0 & 0.101* \\
MVI\_0746 & 23.6 & 0.085 & 6.97 & 739 & 0.26 & 0.47 & 0.001 & 1.64 & 0.07 & 26.30 $\pm \begin{tabular}{@{}c@{}} 3.99 \\ 2.93 \end{tabular}$ & -15.46 & 8.30 $\pm \begin{tabular}{@{}c@{}} 0.99 \\ 1.83 \end{tabular}$ & -11.42 & 0.89 $\pm \begin{tabular}{@{}c@{}} 0.07 \\ 0.06 \end{tabular}$ & 0 & 0.131* \\
MVI\_0748 & 23.6 & 0.068 & 6.97 & 938 & 0.08 & 0.33 & 0.001 & 4.03 & 0.26 & 4.94 $\pm \begin{tabular}{@{}c@{}} 0.72 \\ 0.53 \end{tabular}$ & -20.43 & 2.04 $\pm \begin{tabular}{@{}c@{}} 0.23 \\ 0.42 \end{tabular}$ & -15.98 & 1.01 $\pm \begin{tabular}{@{}c@{}} 0.04 \\ 0.11 \end{tabular}$ & 0 & 0.069 \\
MVI\_0749 & 23.7 & 0.058 & 6.90 & 1362 & 0.08 & 0.02 & 0.000 & 3.59 & 0.08 & 1.84 $\pm \begin{tabular}{@{}c@{}} 0.34 \\ 0.12 \end{tabular}$ & -4.94 & 0.50 $\pm \begin{tabular}{@{}c@{}} 0.08 \\ 0.08 \end{tabular}$ & -19.30 & 1.00 $\pm \begin{tabular}{@{}c@{}} 0.08 \\ 0.05 \end{tabular}$ & 0 & 0.058 \\
MVI\_0750 & 22.8 & 0.060 & 7.58 & 1038 & 0.10 & 0.25 & 0.001 & 1.66 & 0.16 & 7.11 $\pm \begin{tabular}{@{}c@{}} 1.06 \\ 0.69 \end{tabular}$ & -47.60 & 7.43 $\pm \begin{tabular}{@{}c@{}} 1.45 \\ 1.02 \end{tabular}$ & -72.78 & 0.99 $\pm \begin{tabular}{@{}c@{}} 0.07 \\ 0.08 \end{tabular}$ & 0 & 0.038 \\
MVI\_0751 & 23.5 & 0.064 & 7.04 & 1097 & 0 & 0.21 & 0.001 & 1.49 & 0.13 & 5.11 $\pm \begin{tabular}{@{}c@{}} 0.39 \\ 0.64 \end{tabular}$ & -56.11 & 2.01 $\pm \begin{tabular}{@{}c@{}} 0.34 \\ 0.35 \end{tabular}$ & -56.06 & 1.03 $\pm \begin{tabular}{@{}c@{}} 0.06 \\ 0.09 \end{tabular}$ & 0 & 0.051 \\
MVI\_0752 & 23.5 & 0.080 & 7.04 & 920 & 0 & 0.34 & 0.002 & 1.50 & 0.20 & 6.43 $\pm \begin{tabular}{@{}c@{}} 1.10 \\ 0.50 \end{tabular}$ & -56.71 & 4.92 $\pm \begin{tabular}{@{}c@{}} 0.85 \\ 0.77 \end{tabular}$ & -64.81 & 0.96 $\pm \begin{tabular}{@{}c@{}} 0.12 \\ 0.04 \end{tabular}$ & 0 & 0.060 \\
MVI\_0753 & 23.9 & 0.075 & 6.76 & 1068 & 0.10 & 0.23 & 0.001 & 1.94 & 0.08 & 7.79 $\pm \begin{tabular}{@{}c@{}} 1.02 \\ 0.75 \end{tabular}$ & -53.19 & 9.11 $\pm \begin{tabular}{@{}c@{}} 1.44 \\ 1.74 \end{tabular}$ & -70.81 & 1.11 $\pm \begin{tabular}{@{}c@{}} 0.09 \\ 0.08 \end{tabular}$ & -2.73 & 0.016 \\
MVI\_4775 & 29.1 & 0.073 & 4.18 & 1395 & 0 & 0 & 0 & 2.33 & 0.00 & 2.09 $\pm \begin{tabular}{@{}c@{}} 0.36 \\ 0.18 \end{tabular}$ & -43.54 & 0.03 $\pm \begin{tabular}{@{}c@{}} 0.03 \\ 0.03 \end{tabular}$ & 708.16 & 0.99 $\pm \begin{tabular}{@{}c@{}} 0.10 \\ 0.03 \end{tabular}$ & 0 & 0.073 \\
MVI\_4776 & 29.7 & 0.084 & 3.98 & 1391 & 0.15 & 0 & 0 & 5.93 & 0.00 & 3.92 $\pm \begin{tabular}{@{}c@{}} 0.46 \\ 0.45 \end{tabular}$ & -27.89 & 0.10 $\pm \begin{tabular}{@{}c@{}} 0.02 \\ 0.02 \end{tabular}$ & 284.51 & 1.04 $\pm \begin{tabular}{@{}c@{}} 0.04 \\ 0.08 \end{tabular}$ & 0 & 0.084* \\
MVI\_4777 & 29.7 & 0.084 & 3.98 & 1391 & 0.15 & 0 & 0 & 4.37 & 0.00 & 4.19 $\pm \begin{tabular}{@{}c@{}} 0.50 \\ 0.48 \end{tabular}$ & -31.90 & 0.10 $\pm \begin{tabular}{@{}c@{}} 0.02 \\ 0.02 \end{tabular}$ & 284.90 & 1.05 $\pm \begin{tabular}{@{}c@{}} 0.06 \\ 0.07 \end{tabular}$ & 0 & 0.084* \\
MVI\_4778 & 30.2 & 0.088 & 3.82 & 1388 & 0.30 & 0 & 0 & 4.51 & 0.00 & 10.14 $\pm \begin{tabular}{@{}c@{}} 1.13 \\ 0.92 \end{tabular}$ & -9.38 & 1.98 $\pm \begin{tabular}{@{}c@{}} 0.34 \\ 0.28 \end{tabular}$ & -46.07 & 0.94 $\pm \begin{tabular}{@{}c@{}} 0.06 \\ 0.07 \end{tabular}$ & 0.02 & 0.096* \\
\end{tabular}
}}
\caption{Table \ref{tbl:Experiments} cont. }
\end{table}

\clearpage
\begin{table}
\resizebox{\columnwidth}{!}{
\begin{tabular}{l|c|c}
\hline
Source & Relative viscosity & Parameter values\\
\hline
\citet{Einstein1911} & $1 + B\phi$ & $B=1.5-5$ \\
\citet{Roscoe1952} & $(1 - b\phi)^{-B}$ & $b=1-1.35$ \\
Modified Einstein-Roscoe & $(1 - \frac{\phi}{\phi_m})^{-B}$ &  \\
\citet{Krieger1959} & $(1 - \frac{\phi}{\phi_m})^{-B\phi_m}$ & $\phi_m=0.55$ \\
\citet{Mooney1951} & $\exp \left( \frac{B \phi}{1-\phi} \right) $ & \\
\citet{Costa2009} & $\frac{1 + \frac{\phi}{\phi_{*}}^\delta}{(1-F)^{B\phi_{*}}}, F = (1 - \xi) \textrm{ erf}\left(\frac{(\sqrt{\pi}}{2(1-\xi)} \frac{\phi}{\phi_{*}} (1+(\frac{\phi}{\phi_{*}})^\alpha) \right)$
     & $\phi_{*}=0.5, \xi=0.0005, \delta=7, \alpha=5$ \\
\hline
\citet{Taylor1932} & $1 + B\phi$ & $B=1$ \\
\citet{Llewellin2005} & $\eta_{r\inf} + \frac{\eta_{r0} - \eta_{r\inf}}{1 + (K \cdot Ca)^m}, \eta_{r0} = (1-\phi)^{-1}, \eta_{r\inf} = (1-\phi)^{5/3}$ & $K = 6/5, m = 2$ \\
\citet{Princen1989} & $32 (\phi-0.73) Ca^{-1/2}$ & \\
\citet{Stein1992} & $1 + B\phi$ & $B = 13.1$ \\
\hline 
\citet{Phan-Thien1997} & $\begin{cases} (1 - \frac{\phi_{solid}}{1-\phi_{gas}})^{-5/2}(1 - \phi_{gas})^{-1} \\
(1 - \phi_{solid} - \phi_{gas})^{-\frac{5\phi_{solid} + 2\phi_{gas}}{2\phi_{solid} + 2\phi_{gas}}} \\
(1 - \frac{\phi_{gas}}{1-\phi_{solid}})^{-1}(1 - \phi_{solid})^{-5/2} \end{cases}$ & \\
\hline

 & Yield Stress & \\
\hline
\citet{Castruccio2010} & $\begin{cases} 0, & \phi<\phi_c \\ D(\phi - \phi_c)^8, & \phi>\phi_c \end{cases}$ & $\phi_c=0.27$, D=$5\times10^6$ \\
\citet{Mueller2010} & $\tau_*(1 - \phi/\phi_m)^{-2} - 1$ & $\tau_* = \tau(\phi=\phi_m(1 - \sqrt{2}/2)$ \\
\citet{Hoover2001} & $A\left(\frac{\phi/\phi_c-1}{1-\phi/\phi_m}\right)^{1/p}$ & $\phi_c=0.25, \phi_m=0.525, A=5.3, p=1$ \\
\hline
\citet{Princen1989} & $\frac{\sigma \phi^{1/3}}{r}(-0.080 - 0.114\log(1-\phi))$ & \\
\citet{Rouyer2005} & $667(\phi-0.64)^2$ & \\
\hline
 & Flow index & \\
\hline
\citet{Castruccio2010} & $\begin{cases} 1,  & \phi<\phi_c \\ 
C(\frac{\phi_c-\phi}{\phi_m}), & \phi>\phi_c \end{cases}$ & $\phi_c = 0.27, C = 1.3$ \\
\citet{Mueller2010} & $1 - 0.2r_p(\phi/\phi_m)^4$ & \\
\hline
\citet{Truby2015} & $1 - C\phi$ & $C = 0.334$ \\
\hline
\end{tabular}}
\caption{Table of rheology models used in Fig. \ref{fig:lit_rheo} for solid-liquid, gas-liquid, and three-phase models of relative viscosity; solid-liquid and gas-liquid models of yield stress; and solid-liquid, and three-phase models of flow index.}
\label{tbl:rheology_laws}
\end{table}

\end{document}